\documentclass[aps, prd, 12pt, eqsecnum, tightenlines, notitlepage, nofootinbib, floatfix]{revtex4-1}

\PassOptionsToPackage{unicode}{hyperref}

\usepackage{hyperref, graphicx, slashed, xcolor, amsmath}

\allowdisplaybreaks

\definecolor{red}{rgb}{1,0,0}

\newcommand{\beq}{\begin{equation}}
\newcommand{\eeq}{\end{equation}}
\newcommand{\bea}{\begin{eqnarray}}
\newcommand{\eea}{\end{eqnarray}}

\begin{document}

\baselineskip=18pt \pagestyle{plain} \setcounter{page}{1}
\begin{flushright}
CALT-TH-2018-035
\end{flushright}

\vspace*{1.5cm}

\title{The Direct Coupling of Light Quarks to Heavy Di-quarks}

\author{Haipeng An$^{a,b}$ and Mark B. Wise$^c$}
\affiliation{$^a$ Department of Physics, Tsinghua University, Beijing 100084, China}
\affiliation{$^b$ Perimeter Institute for Theoretical Physics, Waterloo, Ontario, N2L 2Y5, Canada}
\affiliation{$^c$ Walter Burke Institute for Theoretical Physics,
California Institute of Technology, Pasadena, CA 91125}

\begin{abstract}

In the limit $m_Q   >m_Q v_{\rm rel}  > m_Q v_{\rm rel}^2  \gg \Lambda_{QCD}$ hadronic states with two heavy quarks $Q$ should be describable by a version of HQET where the heavy quark is replaced by a di-quark degree of freedom. In this limit the di-quark is  a small (compared with $1/\Lambda_{QCD}$) color anti-triplet,  bound primarily by a color Coulomb potential. The excited Coulombic states and  color six states are much heavier than the color anti-triplet ground state. The low lying  spectrum of hadrons containing two heavy quarks is then determined by the coupling of the light  quarks and gluons with momentum of order $\Lambda_{QCD}$ to this ground state di-quark. In this short paper we calculate the coefficient of leading local operator $\left( S_v^{\dagger} S_v\right)  \left({\bar q} {\gamma^{\mu} v_{\mu} } q\right) $ that couples this color-triplet di-quark field  $S_v$  (with  four-velocity $v$) directly to the light quarks $q$ in the low energy effective theory.  It is  ${\cal O}(1/( \alpha_s(m_Q v_{\rm rel}) m_Q^2))$.  While our work is mostly of pedagogical value  we make an  estimate of the contribution of this operator to the masses of $\Xi_{bbq}$ baryon and $T_{QQ{\bar q} {\bar q}}$ tetraquark using the non-relativistic constituent quark model. 
\end{abstract}

\maketitle

\section{Introduction}

The  lowest lying\footnote{We often use a subscript to denote the flavor quantum numbers of a state, particle or field.} $\Xi_{QQq}$ baryons containing two heavy quarks are stable with respect to the strong interactions. For very heavy quarks $Q$  the lowest lying $T_{QQ{\bar q} {\bar q}}$ tetraquark states are also stable with respect to the strong interactions \cite{Carlson,Manohar}.  The reason for this is quite simple. If $m_Q \gg \Lambda_{QCD}$ then, when the heavy di-quark is in a color ${\bar 3}$ configuration, because of the attractive one gluon color Coulombic potential the di-quark  has a large binding energy (compared with $\Lambda_{QCD}$) and a small size (compared with $1/\Lambda_{QCD}$). Strong decay of the lowest lying   $T_{Q Q {\bar q} {\bar q}}$  tetraquark states  to  a baryon with two heavy quarks and an anti-nucleon (when $q=u,d$),  $\Xi_{Q Q q} + {\bar N}_{{q} { q}{ q}}$, is  kinematically forbidden since the final state has  an additional  $q{\bar q}$ pair,  which  costs an additional $\sim 600{~\rm MeV}$ of mass. Strong decay to two heavy mesons $M_{Q {\bar q}} +M_{Q {\bar q}}$ does not require an additional $q {\bar q}$ pair but now the final state does not have the large color Coulombic binding energy proportional to $m_Q$ that the tetraquark state does and so this channel is also kinematically forbidden. 

In nature the heavy quarks that are long lived are the charm and bottom quarks and whether they are heavy enough for tetraquarks containing them to be stable with respect to the strong interactions is not certain but widely believed to be the case  for the lowest lying  tetraquarks with two bottom quarks. See~\cite{Francis:2016hui,Eichten:2017ffp,Karliner:2017qjm} for  recent support for this hypothesis. There are indications that this is not true for the tetraquarks with charm quarks from the previous mentioned studies and~\cite{Cheung:2017tnt}.

Effective field theory methods (discussed mostly in the context of the ${\bar Q} Q$ channel)  have been developed \cite{Brambilla:2004jw,Brambilla:1999xf,Fleming:2005pd}  to take advantage of the fact that for very heavy quarks $Q$ the color anti-triplet  di-quark $QQ$ has a size small compared with $1/ \Lambda_{QCD}$. In this paper we work in the limit $m_Q   >m_Q v_{\rm rel}  > m_Q v_{\rm rel}^2  \gg \Lambda_{QCD}$ where at leading order the light quarks and gluons with momentum of order $\Lambda_{QCD}$  (which we call $\Lambda_{QCD}$ degrees of freedom) in hadrons containing this di-quark  regard the di-quark  as a point object. 

By the sequence of inequalities $m_Q   >m_Q v_{\rm rel}  > m_Q v_{\rm rel}^2  \gg \Lambda_{QCD}$ we mean that while we do treat $v_{\rm rel}$ as small compared with unity, $\Lambda_{QCD}/m_Q$ is much smaller. Hence we will not treat logarithms of the relative velocity as small and resume them. In this case one can match full QCD directly onto an HQET like theory at a scale $\mu$ which we take to be $\mu=m_Q v_{\rm rel}$.

In this limit the color six $QQ$ configurations and Coulombic excitations above the lowest lying color anti-triplet di-quark state (described by principal quantum numbers $n>1$) are  much heavier than the ground state and can be integrated out of the theory. Hence (in the single di-quark sector) one arrives at a theory like HQET \cite{Georgi,Eichten} with Lagrange density,

\begin{equation}
\label{leading}
{\cal L}= S^{\dagger}_{v} iv^\mu D_\mu S_v  -{1 \over 4} G^{A \mu \nu}G^A_{\mu \nu} +\sum_q {\bar q } (i  {\gamma^{\mu} D_{\mu}} -m_q)q +{ \ldots}
\end{equation}
 Including the heavy quark fields\footnote{After including the heavy quark fields $Q$ the effective field theory is used in the one heavy quark  and one heavy di-quark sectors.} the leading terms would not only have the familar heavy quark spin-flavor symmetry  \cite{Isgur} but also an enlarged heavy quark and di-quark spin-flavor symmetry~\cite{Savage:1990di,Georgi:1990ak,Eichten:2017ffp,Hu:2005gf,Mehen:2017nrh}.

$\Lambda_{QCD}$ gluons coupling directly to the heavy di-quark already occur at leading order through the covariant derivative $ D=\partial -i g {\bar T}^B A^B$ where the bar denotes that the SU(3) generators are in the ${\bar 3} $ representation.
Of course the $\Lambda_{QCD}$  quarks $q$ interact with the gluons, so even at leading order they interact with the di-quark. The purpose of this paper is to calculate the direct coupling of the $\Lambda_{QCD}$ quarks $q$ to the di-quark field $S$ which occurs in the ellipses of eq.~(\ref{leading}). We find that this operator is of the form, $\left( S_v^{\dagger} S_v\right)  \left({\bar q} {\gamma^{\mu} v_{\mu} } q\right) $, and that its coefficient occurs at order ${\cal O}(1/( \alpha_s(m_Q v_{\rm rel}) m_Q^2))$ which is between ${\cal O}(1/m_Q)$ and  ${\cal O}(1/m_Q^2)$. The $1/( \alpha_s(m_Q v_{\rm rel})) $ arises because this term is suppressed by the di-quark size\footnote{Direct couplings of the light quarks with momentum of order the QCD scale to a heavy di-quark were also considered in~\cite{Fleming:2005pd}. However, the effects they focussed on don't arise from the finite size of the di-quark and are smaller than those we consider  in this paper.}. We find it interesting to compute the coefficient of this term because it gives rise to dependence on the heavy quark mass and hence a breaking of heavy quark  di-quark flavor symmetry that arises from the size of the di-quark system. The pattern and heavy quark mass dependence of its contribution to the breaking of heavy quark di-quark flavor symmetry is different from that of the heavy quark and di-quark kinetic terms that arise at ${\cal O}(1/m_{Q})$ (i.e.,  $\Delta L_{\rm kin} ={\bar Q_v} D^2 Q_v/(2 m_Q) +S_v^{\dagger} D^2 S_v/(2 m_S)$). For example, the term we are focussing on  contributes to the  $\Xi_{QQq}$ baryon mass but not to the $M_{ Q {\bar q}}$ meson mass, while  $\Delta L_{\rm kin}$ contributes to both. 

In nature the heavy quarks are the top, bottom and charm quarks. While the top is very heavy compared with the QCD scale, it is short lived and does not form hadronic bound states. That leaves the bottom and charm quarks. Dimensional analysis suggests that for neither of these quarks will  approximations based on $ m_{b,c} v_{\rm rel}^2 \gg \Lambda_{QCD}$ be valid and even predictions based on the condition $m_{b,c} v_{\rm rel}\gg \Lambda_{QCD}$ are suspect although it is likely that in the bottom quark case that they have some utility. Hence we view our work as mostly of pedagogical value.  Despite these cautionary remarks we will make an  estimate of the importance  of the operator $\left( S_v^{\dagger} S_v\right)  \left({\bar q} {\gamma^{\mu} v_{\mu} } q\right) $ to the masses of the lowest lying $\Xi_{bbq}$ baryon and $T_{Q Q {\bar q} {\bar q}}$ tetraquark using the non-relativistic constituent quark model.  

\section{The ground state color ${\bar 3}$ di-quark}

Di-quarks  can be in either a color triplet or color six  representation. The $3 \times 3 \rightarrow {\bar 3}$ channel is attractive and after tracing over the color the short range color Coulombic potential is,
\begin{equation}
V(r) =- {2\over 3}{\alpha_s \over r}.
\end{equation}
This is half as strong as the attractive potential  in the  $3 \times {\bar 3} \rightarrow 1$ channel, which is relevant for quarkonium. While it is a bit odd to consider the qualitative reason for a factor of two difference  between the potentials in these two channels, one can be found in the large number of colors $N_c$ limit~\cite{tHooft}
.  Assuming  (just for simplicity) that all the quark flavors are different in the large $N_c$ limit the interpolating field for  a $\Xi_{Q_1Q_2 q_3, \ldots q_{N_c-2}}$ baryon is (suppressing constants, and all indices except for flavor and color) $\epsilon_{\beta_1 \beta_2 \alpha_1 \ldots   \alpha_{N_c-2}} Q_{1\beta_1} Q_{2 \beta_2}q_{1\alpha_1} \ldots q_{N_c-2\alpha_{N_c-2}}$ and the short range effective potential between the two heavy quarks in this state is,

\begin{equation}
\label{potential}
V(r) =- {N_c+ 1 \over 2 N_c}{\alpha_s \over r}.
\end{equation}
This is suppressed by a factor of $1/N_c$ in the large number of colors limit where $N_c \alpha_s$ is held fixed as $N_c \rightarrow \infty$. At large $N_c$ the appropriate interpolating field for tetraquarks with two heavy quarks $T_{Q_1Q_2 {\bar q}_1{\bar q}_2}$  is $\left(Q_{1 \alpha} Q_{2 \beta}{\bar q}_{1 \alpha} {\bar q}_{2 \beta} -Q_{1 \beta} Q_{2 \alpha}{\bar q}_{1 \alpha} {\bar q}_{2 \beta} \right)$ and eq.~(\ref{potential}) still applies for the color Coulombic potential between the heavy quarks. On the other hand, for the case of $ {Q \bar Q}$ quarkonium with $N_c$ colors the appropriate interpolating field is ${\bar Q}_{\alpha}Q_{\alpha}$ and the color Coulombic potential is

\begin{equation}
V(r) =- {N_c^2-1 \over 2 N_c}{\alpha_s \over r},
\end{equation}
which does does not vanish in the large $N_c$ limit.

Returning to the real world where $N_c=3$ the color ${\bar 3}$ di-quark states are twice the size of the color singlet quarkonium states and so the multipole expansion should be somewhat less reliable in the di-quark case.

The ground state  di-quark has a spatial wave-function
\begin{equation}
\phi({\bf r})= \left ( {1 \over \pi a_0^3  }       \right)^{1/2} e^{-r/a_0},
\end{equation}
where the Bohr radius is
\begin{equation}
a_0= {3 \over 2 \alpha_s  \mu_Q},
\end{equation}
and the reduced heavy quark mass is $\mu_Q=m_{Q_1}m_{Q_2}/(m_{Q_1}+m_{Q_2})$. In the case where the heavy quarks are the same flavor they must be in a spin-one state. When they are different there are degenerate spin-zero and spin-one cases. Spin is usually inert for our purposes in this paper and we will usually not keep track of those labels in our equations.  

Later we will need the square of the charge radius,

\begin{equation}
\langle r^2 \rangle=\int d^3r r^2 |\phi({\bf r})|^2= 3 a_0^2= {27 \over 4 \alpha_s^2 \mu_Q^2} .
\end{equation}

Since were are not treating $v_{\rm rel}$ as very small the argument of the strong coupling can be taken to be either $m_Q$ or $m_Q v_{\rm rel}$, in the equations given in this section.  However the latter is physically more appropriate so  we will use it for any quantitative estimates we make using these formulae going forward.

\begin{figure}
\centering
\includegraphics[height=1.3in]{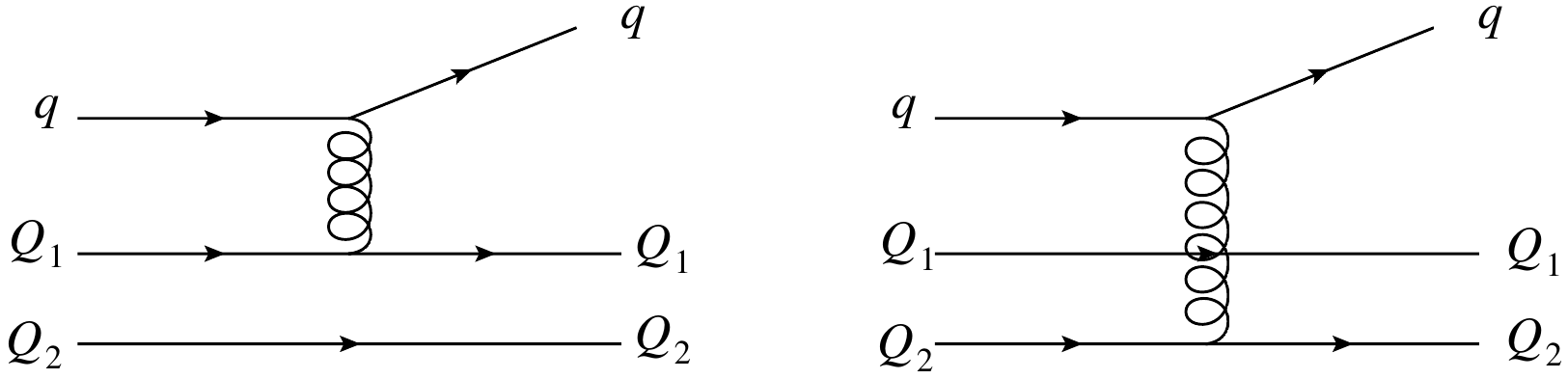}
\caption{Feynman diagrams that contribute (at tree level) to elastic light quark heavy di-quark scattering.}\label{fig:1}
\end{figure}

\section{Matching}

One can compute  matching onto the effective HQET like di-quark effective field theory by computing an appropriate physical perturbative process in QCD. We will work in the leading logarithmic approximation, which means tree level matching and one-loop renormalization group running. By physical we mean on-shell but not  taking into account confinement. Since we are interested in local operators involving the light quarks and the di-quark field $S$ the appropriate process is light quark heavy di-quark elastic scattering, $q_{\beta}({k_i})+ (Q_1Q_2)_{\alpha} \rightarrow (Q_1Q_2)_{\alpha'}+ q_{\beta'}(k_f)$, as shown in Fig.~\ref{fig:1}. In the heavy quark limit the scattering must be elastic,  $k_f^0=k_i^0$  and we denote the three-momentum transfer by ${\bf k}={\bf k}_i-{\bf k}_f$ and $k=|{\bf k}| $. We work in the rest frame of the di-quark.

Expanding the heavy quark spinors to zeroth order in three-momenta we  find that the amplitude ${\cal A}$ for this process is
\begin{eqnarray}
\label{physical}
&&{\cal A}\simeq {g^2 \over 2 k^2}\left(2 m_{Q_1} +2 m_{Q_2} \right){\bar u}(k_f) T^A_{\beta' \beta}\gamma^0 u(k_i) \left( -T^A  \right)_{\alpha \alpha'} \times \nonumber \\
&& \int {d^3 p \over (2 \pi)^3}\left( {\tilde \phi}^* \left({\bf p}-{m_{Q_2} \over m_{Q_1}+m_{Q_2} } {\bf k}  \right)+ {\tilde \phi}^* \left({\bf p}+{m_{Q_1} \over m_{Q_1}+m_{Q_2} } {\bf k}  \right)\right)\tilde\phi({\bf p}).
\end{eqnarray}
Here $\tilde \phi$ is the Fourier transform of the spatial wave-function for the di-quark state.  Expanding eq.~(\ref{physical}) in $\bf k$ the term at zero'th order corresponds to the leading order Lagrangian in eq.~(\ref{leading}) where the light quark scatters off the anti-triplet charge of the di-quark without resolving its size. The term linear order in $\bf k$ vanishes because the ground state wave-function $\phi$ is s-wave. The term quadratic order in $k$ is,
\begin{equation}
{\cal A}_2 \simeq { g^2 \over 2} \left(2 m_{Q_1} +2 m_{Q_2} \right){\bar u}(k_f) T^A_{\beta' \beta}\gamma^0 u(k_i) \left( -T^A  \right)_{\alpha \alpha'} \left(-{\langle r^2 \rangle \over 6} \right) {m_{Q_1}^2+m_{Q_2}^2 \over ({m_{Q_1}+m_{Q_2}})^2 },
\end{equation}
where the subscript 2 denotes that we have expanded to quadratic order in $k$.

Matching onto the effective theory we find the contribution (generalizing to arbitrary di-quark four-velocity $v$) to its Lagrange density
\begin{equation}
\Delta {\cal L}=  C \left( S_v^{\dagger} {\bar T}^A S_v \right) \sum_q  \left({\bar q} T^A \gamma^{\mu}v_{\mu} q \right)
\end{equation}
where ${\bar T}^A=-(T^A)^T$ are the $SU(3)$ generators in the anti-triplet representation and the coefficient
\begin{equation}
\label{match}
C= {\pi \alpha_s \langle r^2 \rangle \over 3}  \left( {m_{Q_1}^2+m_{Q_2}^2 \over ({m_{Q_1}+m_{Q_2}})^2 } \right)= {9 \pi \over 4 \alpha_s}\left ( {m_{Q_1}^2+m_{Q_2}^2 \over  m_{Q_1}^2 m_{Q_2}^2} \right)
\end{equation}
The  charge radius arises from the non-zero size of the di-quark  as so we feel it is appropriate to view the coefficient in eq.~(\ref {match}) as evaluated at the subtraction point $\mu =m_Q v_{\rm rel}$ even though as mentioned earlier we will not be keeping track of factors of $v_{\rm rel}$ in logarithms as far as the power counting is concerned. 

The cancellation of the gluon propagator's $1/k^2$ by the factor of $k^2$ from expanding the wave-functions has the same origin as in penguin diagrams for weak decays and can be thought of as arising from an application of the equations of motion \cite{Wise}.
Finally we remove the product of SU(3) generators using the identity (recall ${\bar T}^A=(-T^{A})^T$), $T_{\alpha \beta}^AT_{\mu \nu}^A={-(1/6)}\delta_{\alpha \beta}\delta_{\mu \nu}+{(1/2)} \delta_{\alpha \nu}\delta_{\beta \mu}$, and write the effective Lagrangian as
\begin{equation}
\Delta {\cal L}=  C_1 O_1+C_2 O_2
\end{equation}
where
\begin{equation}
O_1=\left(S^{\dagger}_{v \alpha} S_{v \alpha}\right) \sum_q \left({\bar q }_{\beta} \gamma^{\mu}v_{\mu} q_\beta \right) ,~~ O_2=\left( S^{\dagger}_{v \alpha} S_{v \beta} \right)\sum_q \left({\bar q }_{\alpha} \gamma^{\mu}v_{\mu} q_\beta \right)
\end{equation}
 and
  \begin{equation}
  \label{initial values}
  C_1 = C/6, ~~ C_2= -C/2.
  \end{equation}
\section{Running}

\begin{figure}
\centering
\includegraphics[height=2in]{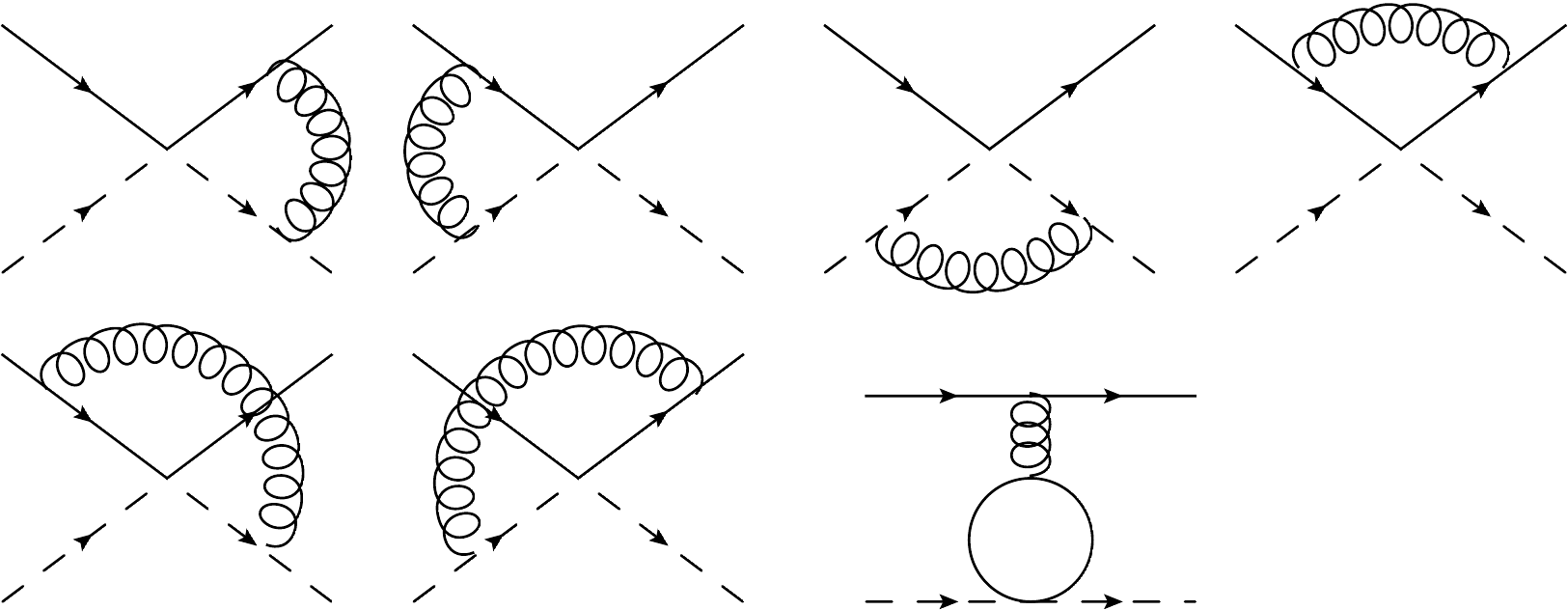}
\caption{Feynman diagrams that contribute to the anomalous dimension matrix for the operators $O_{1,2}$. The last is not one-particle irreducible but contributes to a local operator in the same way the penguin diagrams do in kaon decay.}\label{fig:2}
\end{figure}

Although we are not keeping track of logarithms of $v_{\rm rel}$ we do want to sum logs of the ratio $\Lambda_{\rm QCD}/m_Q$ using the renormalization group. The values of the coefficients $C_{1,2}$ in eq.~(\ref{initial values}) are interpreted as evaluated at a subtraction point $\mu \sim m_Q v_{\rm rel}$. To scale down to a lower value of the subtraction point, we need the anomalous dimension matrix for the operators $O_{1,2}$ calculated in the leading order effective  Lagrange density displayed explicitly in eq.~(\ref{leading}). The subtraction point dependence of the operators $O_{1,2}$ is given by the renormalization group equations,
\begin{equation}
\mu{d \over d \mu} O_j=-\gamma_{ji} O_i.
\end{equation}
This anomalous dimension matrix $\gamma$ is computed from the one loop diagrams 
Feynman diagrams in Fig.~\ref{fig:2}.  Using dimensional regularization with minimal subtraction  we find that,

\begin{equation}
\gamma(g)= {g^2 \over 16 \pi^2}
\begin{pmatrix}
0 & 0 \\
3-{4 \over 9}n_q & -9+{4 \over 3}n_q
\end{pmatrix},
\end{equation}
where $n_q$ is the number of light quark flavors.

A basis of operators that are multiplicatively renormalized are $O_1$ and $O_- = O_1-3O_2$ which is the linear combination of $ O_{1,2}$
that we matched onto at the scale $m_Q v_{\rm rel}$. The operator $O_-$ has anomalous dimension,
\begin{equation}
\gamma_-(g)={g^2 \over 16 \pi^2} \left(-9+{4 \over 3}n_q \right ).
\end{equation}
 Combining this with the results of the  previous section we arrive at
 \begin{equation}
 \label{final result}
 \Delta L=  \left({3 \pi \over 8 \alpha_s (m_Q v)}\right)\left ( {m_{Q_2}^2+m_{Q_1}^2 \over  m_{Q_2}^2 m_{Q_1}^2} \right)\left[ {\alpha_s(m_Q v) \over \alpha_s(\mu)} \right]^{\left(-{9 \over 2} +{2 \over 3} n_q \over 11-{2 \over 3} n_q \right)} O_-(\mu) 
\end{equation}
where
\begin{equation}
O_-=\left(S^{\dagger}_{v \alpha} S_{v \alpha}\right) \sum_q \left({\bar q }_{\beta} \gamma^{\mu}v_{\mu} q_\beta \right) -3\left( S^{\dagger}_{v \alpha} S_{v \beta} \right)\sum_q \left({\bar q }_{\alpha} \gamma^{\mu}v_{\mu} q_\beta \right).
\end{equation}
The two equations above are the main results of this paper.

Including logarithmic corrections to scale the effective Lagrangian down from the scale $m_Q v_{\rm rel}$ is not just of academic interest.  As an example of how it can matter consider the color magnetic moment term that arises in the matching from  expanding the heavy quark spinors to leading order in the gluon momentum. It gives rise in the rest frame of the di-quark  (when the two heavy quarks composing the di-quark are identical) to the term\footnote{In this case the anomalous scaling is the same as in the heavy quark case \cite{EichtenHill,Falk}.}.
\begin{equation}
\Delta L=  {1 \over 4 m_Q} \left[ {\alpha_s( m_Q v_{\rm rel} )  \over  \alpha_s( \mu)}\right] ^{\left( {9 \over 33-2 n_q} \right)} \left( S^{\dagger} {\bar T}^A {\bf S}   S \right) g {\bf B}_{\rm color}^A
\end{equation}
where $ {\bf S}$  is the di-quark spin vector. If we had not evaluated the strong coupling at $\mu$ but rather at the matching scale  the anomalous dimension of the operator would be large and effectively bring the scale the coupling is evaluated at  to $\mu$.  What we have seen in this section is that a large anomalous scaling  like this does not occur for the local operator $O_-$.

\section{A Non-relativistic Constituent Quark Model Estimate}

We can get a rough idea about how large the contribution of eq.~(\ref{final result})  is to the mass of the $\Xi_{bbq}$ baryon by making a non-relativistic quark model estimate of the matrix element of $O_- $ which presumably we should view as reasonable for a subtraction point $\mu$ around the QCD scale.  In the non relativistic constituent quark model color is included through a color factor and then the light quarks are viewed as non-relativistic quasi-particles bound by some potential. By relating various physical quantities in the model an estimate can be made independent of the particular potential that  binds the constituent quarks in a hadron. The estimate we make in this section is similar in spirit to using the vacuum insertion approximation \cite{Gaillard} for the $K- {\bar K}$ matrix element of the four-quark operator $({\bar s} \gamma^{\mu}(1- \gamma_5) d)( {\bar s}\gamma_{\mu}(1- \gamma_5) d)$.

The color configuration for the $\Xi_{bbq}$ is $ (1/ \sqrt{3}) S_{\alpha}q_{\alpha}$. In the non-relativistic quark model, for a $\Xi_{bbq}$ at rest, we find the $O_-$ expectation value to be
\begin{equation}
\left\langle  \int d^3 x O_- ({\bf x})  \right\rangle_{\Xi}  =-8 \int d^3x ~ n_S({\bf x}) n_q({\bf x}) =-8 |\phi_q({\bf 0})|^2,
\end{equation}
where  $n_S({\bf x}) =\delta^3({\bf x})$ is the number density of di-quarks,  $n_q({\bf x})= |\phi_q({\bf x})|^2$ is the number density of light quarks $q$, $\phi_q$ is the wave function of $q$, and $-8$ is the color factor. Using heavy quark symmetry $\phi({\bf 0})$ is related to the $B$-meson decay constant,
\begin{equation}
\phi({\bf 0})= {f_B \sqrt{m_B} \over {2\sqrt{3}} }\left[ {\alpha_s(m_b) \over \alpha_s(\mu)}\right]^{6 \over 33-2 n_q}.
\end{equation}
Combining these results, neglecting the renormalization group running and setting $m_b=m_B$, we have that the contribution of the Lagrange density in eq.~(\ref{final result}) to the $ \Xi_{bbq}$ mass, $\Delta m_{\Xi_{bbq}}$, is estimated to be
\begin{equation}
\label{number}
\Delta m_{\Xi_{bbq}} \simeq {\pi \over 2\alpha_s(m_b v)}{ f_B^2 \over m_B} \sim 30~{\rm MeV}.
\end{equation}
Here we used $f_B \simeq 190~ {\rm MeV}$ and   $\alpha_s(m_b v) \simeq 0.35$ for the  numerical result. The  numerical result in eq.~(\ref{number}) above is only a little smaller than a typical order $\Lambda_{\rm QCD}^2/m_b$ contribution to the $\Xi_{bbq}$ mass. This should not be particularly surprising given that the Bohr radius for such a color Coulombic bound state is $a_0= 3/(\alpha_s(m_bv) m_b) \sim 1/(600~{\rm MeV})$. 

The color configuration for the $T_{bb{\bar q}{ \bar q}}$ tetraquark is $(1/\sqrt{6}) \epsilon^{\alpha \beta \gamma} S_{\alpha} \bar q_{\beta} \bar q_{\gamma}$ (there are spin flavor labels on the light quarks that we have suppressed). 
In the non-relativistic constituent quark model, for a $T_{bb{\bar q}{\bar q} }$ at rest  expectation value, we find that
\begin{equation}
\left\langle  \int d^3 x O_- ({\bf x})  \right\rangle_{T}  = 4 \int d^3x ~ n_S({\bf x}) (-n_{\bar q}({\bf x})) = -4 n_{\bar q}({\bf 0}).
\end{equation}
The color factor is $-1/2$ what it was for the baryon case and  there is an additional minus sign because ${\bar q} \gamma^0 q =n_q-n_{\bar q}$.  Since  $T_{bb{\bar q}{\bar q}}$ contains two antiquarks (while $\Xi_{bbq}$ contains a single quark)  a contribution of about $30~{\rm MeV}$ to the mass of the $T_{bbqq}$ tetraquark from this term is a reasonable estimate. Note that if the contribution of $O_-$ to the mass of the  $T_{QQqq}$ tetraquark and $\Xi_{QQq}$ baryon are the same then  $\Delta L$ in eq.~(\ref{final result} ) does not correct the leading order sum rule~\cite{Eichten:2017ffp}, $m_{QQ{\bar q}{\bar q}}-m_{Qqq}=m_{QQq}-m_{Q{\bar q}}$.

Of course there are additional contributions to the masses of the $\Xi_{bbq}$ and $T_{bb{\bar q}{\bar q}}$ hadrons  from the leading terms explicitly displayed in eq.~(\ref{leading}) and the familiar (from HQET) terms of order $1/m_{Q}$. However these do not arise from the size of the heavy di-quark and have a  different pattern of contributions to the masses of hadrons containing one heavy quark or di-quark and a different dependence on the heavy quark mass.

\section{Why do we  take $m_Qv_{\rm rel}^2 \gg \Lambda_{QCD}$ }

This paper is about the effective field theory for the ground state anti-triplet di-quark and the direct coupling of light $\Lambda_{QCD}$ quarks to the ground state di-quark degrees of freedom in that effective HQET like theory. If we did not  take\footnote{Including numerical factors in the ground state di-quark color Coulombic binding energy we need $ \alpha_s(m_Q v)^2m_Q  /9 \gg \Lambda_{QCD} $ when the two heavy quarks are the same.} $m_Qv_{\rm rel}^2 \gg \Lambda_{QCD}$ then such an effective field theory would not be appropriate. One could still write an effective theory~\cite{Hu:2005gf,Mehen:2017nrh} for the lowest lying baryons (or tetraquarks) containing two heavy quarks interacting with low momentum photons and pions, or an effective theory containing the possible di-quark configurations (pNRQCD) but matching the latter to an effective field theory just containing the lowest lying di-quark configuration  and the $\Lambda_{QCD}$  gluon and light quark degrees of freedom would not be justified. 

To illustrate this let us consider the case where the two heavy quarks are different flavors. Then expanding eq.~(\ref{physical}) to linear order in ${\bf k}$ we match onto a transition operator taking the lowest lying ($n$=2) $L=1$ color anti=triplet di-quark \footnote{We take the two heavy quarks to be in the spin-zero configuration so the total spin of the initial  di-quark is one and the final di-quark is zero.} $S_j$  to the lowest lying ($n$=1) $L=0$  di-quark field $S$ we have been considering. In the rest frame of the di-quarks, 
\begin{equation}
\label{higherorder}
\Delta L ={1 \over 2 {\sqrt 3}} \left( { m_{Q_2}-   m_{Q_1} \over    m_{Q_2} + m_{Q_1}}  \right)S^{\dagger }_j {\bar T}^ASgE_{\rm color}^{A j}  \langle r \rangle_{\rm trans} +{\rm h.c.}
\end{equation}
where the transition charge radius is
\begin{equation}
\langle r \rangle_{\rm trans}=\int_0^{\infty} dr r^3 R_{2,1}(r) R_{1,0}(r)= {\sqrt {2 \over 3}} \left({128 \over 81} \right)  a_0
\end{equation}
eq.~(\ref{higherorder}) contributes to the mass of a $\Xi_{Q_1Q_2 q}$ baryon at second order in $\Delta L$  an amount of  order $\Delta m_{\Xi_{Q_1Q_2 q}} \sim \Lambda_{QCD}^4/(\alpha_s (m_Q v_{\rm rel})^4 m_Q^3)$.  Here the strong coupling $g$ in eq.~(\ref{higherorder}) is evaluated at the subtraction point  (i.e., near the QCD scale) and not the matching scale since we know from HQET that there is a large anomalous dimension that  makes this appropriate. Recall that  the contribution  from the matrix element of $O_{-}$ estimated in the previous section (see eq.~(\ref{number}))  is 
of order $\Delta m_{\Xi_{Q_1Q_2 q}} \sim \Lambda_{QCD}^3/(\alpha_s (m_Q  v_{\rm rel}) m_Q^2)$. So the impact on the $\Xi_{Q_1Q_2q}$ mass from eq.~(\ref{higherorder}) at second order in perturbation theory is suppressed by a factor of $ \Lambda_{QCD}/(\alpha_s (m_Q  v_{\rm rel})^3 m_Q)=(1/ v_{\rm rel})\times(\Lambda_{QCD}/(m_Qv_{\rm rel}^2))$ when compared with the contribution of $O_-$. This contribution and the contribution of other excited di-quark states (including the color six continuum and color anti-triplet scattering states) would not be suppressed if we did not work in the limit  $m_Qv_{\rm rel}^2 \gg \Lambda_{QCD}$.
\section{Concluding Remarks}

In this paper we have computed the leading direct coupling of the quarks that have momenta of order $\Lambda_{QCD}$ to the effective color anti-triplet di-quark degree of freedom $S$ assuming the hierarchy of scales,  $m_Q   >m_Q v_{\rm rel}  > m_Q v_{\rm rel}^2  \gg \Lambda_{QCD}$. In the effective ${\rm HQET}$ like theory for di-quarks this comes from the  operator $\left( S_v^{\dagger} S_v\right)  \left({\bar q} {\gamma^{\mu} v_{\mu} } q\right) $ which corresponds in the baryon $\Xi_{QQq}$ to a repulsive delta function potential between the heavy di-quark and the light quarks and in the $T_{QQ {\bar q} {\bar q}}$  a repulsive delta function potential between the heavy di-quark and the light anti-quarks. It arises from the finite size of the di-quark and has a coefficient  ${\cal O}(1/( \alpha_s(m_Q v_{\rm rel}) m_Q^2))$. Its coefficient is anomalously large because the factor of $1/\alpha_s(m_Q v_{\rm rel})$ originates from $ g(m_Q v_{\rm rel})^2/\alpha_s(m_Q v_{\rm rel})^2$ which gives an additional $4 \pi$ when written in terms of color fine structure constant\footnote{Hence higher order terms in the multipole expansion that arise from expanding eq.~(\ref{physical}) to higher orders in ${\bf k}$ will not be even more enhanced.}. We estimated, using the non-relativistic  quark model, that this term would contribute around $30~{\rm MeV}$ to the mass of tetraquarks and baryons containing two bottom quarks.  It gives rise to  the leading violation of heavy quark, di-quark flavor symmetry arising from the finite size of the di-quark. 

If the stability (with respect to the strong interactions) of tetraquarks containing two heavy bottom quarks is firmly established then it will still be interesting to study other aspects of their physical properties.  For example, will they correspond more to the small (compared with $1/\Lambda_{QCD}$) di-quark picture or to a di-meson molecule. The latter is possible since the long range potential from one pion exchange is attractive in some channels and capable of giving rise to bound states~\cite{Manohar}. Perhaps tetraquarks  that  contain two heavy bottom quarks and are stable with respect to the strong interactions will lie between these two extremes.

\acknowledgments
We thank Aneesh Manohar, Tom Mehen, Ira Rothstein, and Mikhail Solon for some useful comments. HA is supported by the Recruitment Program for Young Professionals of the 1000 Talented Plan and the Tsinghua University Initiative Scientific Research Program. MBW is supported by the DOE Grant DE-SC0011632 and the Walter Burke Institute for Theoretical Physics. Research at Perimeter Institute is supported by the Government of Canada through Industry Canada and by the Province of Ontario through the Ministry of Economic Development \& Innovation.

\end{document}